\documentclass[11pt,twoside]{article}  
  
\usepackage{asp2006}  
\usepackage{epsf}  
\usepackage{psfig}  
\usepackage{lscape}  
\usepackage{graphicx} 
  
\begin{document}  
\title{WELS - Ultraviolet Spectra and Expanding Atmosphere Models}   
\author{W. L. F. Marcolino}  
\affil{Laboratoire d'Astrophysique de Marseille, France}   
  
\begin{abstract} 
The ultraviolet spectra of all "weak emission line central stars of planetary 
nebulae" (WELS) with available IUE data is analyzed. We found that the WELS  
can be divided in three different groups regarding their UV:  
(1) Strong P-Cygni profiles (mainly in C IV 1549); (2) Weak P-Cygni features 
and (3) Absence of P-Cygni profiles. We have measured wind terminal velocities for  
all objects presenting P-Cygni profiles in N V 1238 and/or C IV 1549. The  
results obtained were compared to the UV data of the two prototype stars of 
the [WC]-PG 1159 class, namely, A30 and A78. They indicate that WELS are 
distinct from the [WC]-PG 1159 stars, in contrast to previous claims in the  
literature. In order to gain a better understanding about the WELS, we clearly 
need to determine their physical parameters and chemical abundances. First non 
LTE expanding atmosphere models (using the CMFGEN code) for the UV and optical 
spectra of the star Hen 2-12 are presented. 
\end{abstract}  
 
\section{Introduction:} 
  
Hydrogen deficient central stars of planetary nebulae (CSPN)  
are generally divided in three main groups\footnote{Here, we neglect 
other type of hydrogen deficient objects such as R Coronae Borealis (RCB)  
and O(He) stars}: [WR], PG 1159 and [WC]-PG 1159 stars.  
However, besides these classes, there is the "weak emission  
line stars" (WELS or [WELS]) group, which was introduced in the  
extensive observational study presented by Tylenda et al. (1993).  
 
Although a considerably advance in the understanding of the origin  
and evolution of [WR], [WC]-PG 1159 and PG 1159 stars have been  
achieved in the last decade (see e.g. Werner \& Herwig 2006),  
the evolutionary status of the WELS remains an open question.  
Motivated by the fact that only a few studies have been  
done focusing these objects, and that most of them were done in the  
optical part of the spectrum, we have investigated the UV spectra  
of all WELS with available IUE data. Moreover, we have applied  
the non LTE expanding atmosphere code CMFGEN to model the UV  
and optical spectra of one WELS, namely, Hen 2-12.  
 
\section{The UV Spectra of the WELS:} 
 
\begin{figure}[!ht] 
\centering 
\includegraphics[width=270pt,height=220pt]{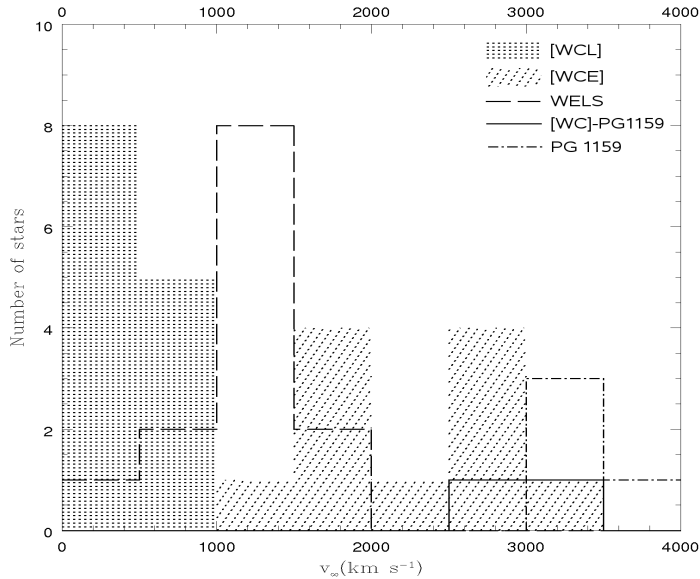} 
\caption{Terminal wind velocities of hydrogen deficient CSPN (from Marcolino et al. 2007).} 
\label{vinfres} 
\end{figure} 
 
We have used the MAST database to retrieve all the data available for  
the WELS from the IUE satellite. The total number of objects currently known  
is about 50, but only about 40\% of this population was observed by the IUE.  
For all the details regarding the UV analysis, we refer the
reader to the complete results published in Marcolino et al. (2007). 
 
Instead of presenting a homogeneous set of features, we found  
that the WELS could be divided in three different groups:  
1) Strong P-Cygni profiles (mainly in C IV 1549); (2) Weak P-Cygni  
feature in C IV 1549 and (3) Absence of P-Cygni profiles.  
In the case of group (3), the lines are very intense and 
most likely of nebular origin. 
 
We have compared the UV spectra of the WELS with the UV  
spectra of the two prototypes of the [WC]-PG 1159 class: A30 and A78. 
These two objects present simultaneous intense P-Cygni emissions  
of N V 1238, O V 1371 and C IV 1549. In contrast, in the WELS, the  
O V 1371 line is weak or absent. The same is true for N V 1238  
in some objects. It is also conspicuous that the P-Cygni profiles  
in A30 and A78 present a broader absorption than in the WELS spectra. 
These characteristics indicate that the WELS are not  
[WC]-PG 1159 stars, in contrast with previous claims in the literature 
(see Parthasarathy et al. 1998) on the basis of optical spectra.
However, the situation is not clear for the stars NGC 6543, NGC 6567, and 
NGC 6572. They present simultaneous P-Cygni emissions of N V 1238, 
O V 1371 and C IV 1549, but considerably lower terminal velocities than 
the ones found for A30 and A78 (see Marcolino et al. 2007 for more details).

\subsection{Wind Terminal Velocities:} 
 
After the spectral comparison between the WELS and the [WC]-PG 1159  
stars A30 and A78, we have derived wind terminal velocities ($v_\infty$) for all  
objects presenting P-Cygni profiles in N V 1238 and/or C IV 1549.  
As most of the data available are of low resolution, we have used  
the calibration provided by Prinja (1994). For spectra of high resolution, 
the standard procedure to measure $v_\infty$ was used. 
Our results are displayed in Fig. \ref{vinfres}, along with data for  
[WCE], [WCL] and PG 1159 stars, obtained from  
Koesterke (2001). As it is clear from the figure, the $v_\infty$ distribution 
for the WELS stand mainly between the early and late type [WR] stars  
distribution. Moreover, the WELS have much lower terminal velocities  
than the [WC]-PG 1159 and PG 1159 stars.  
 
\section{First models for Hen 2-12:}

\begin{figure}[!t] 
\centering 
\includegraphics[width=430pt,height=430pt]{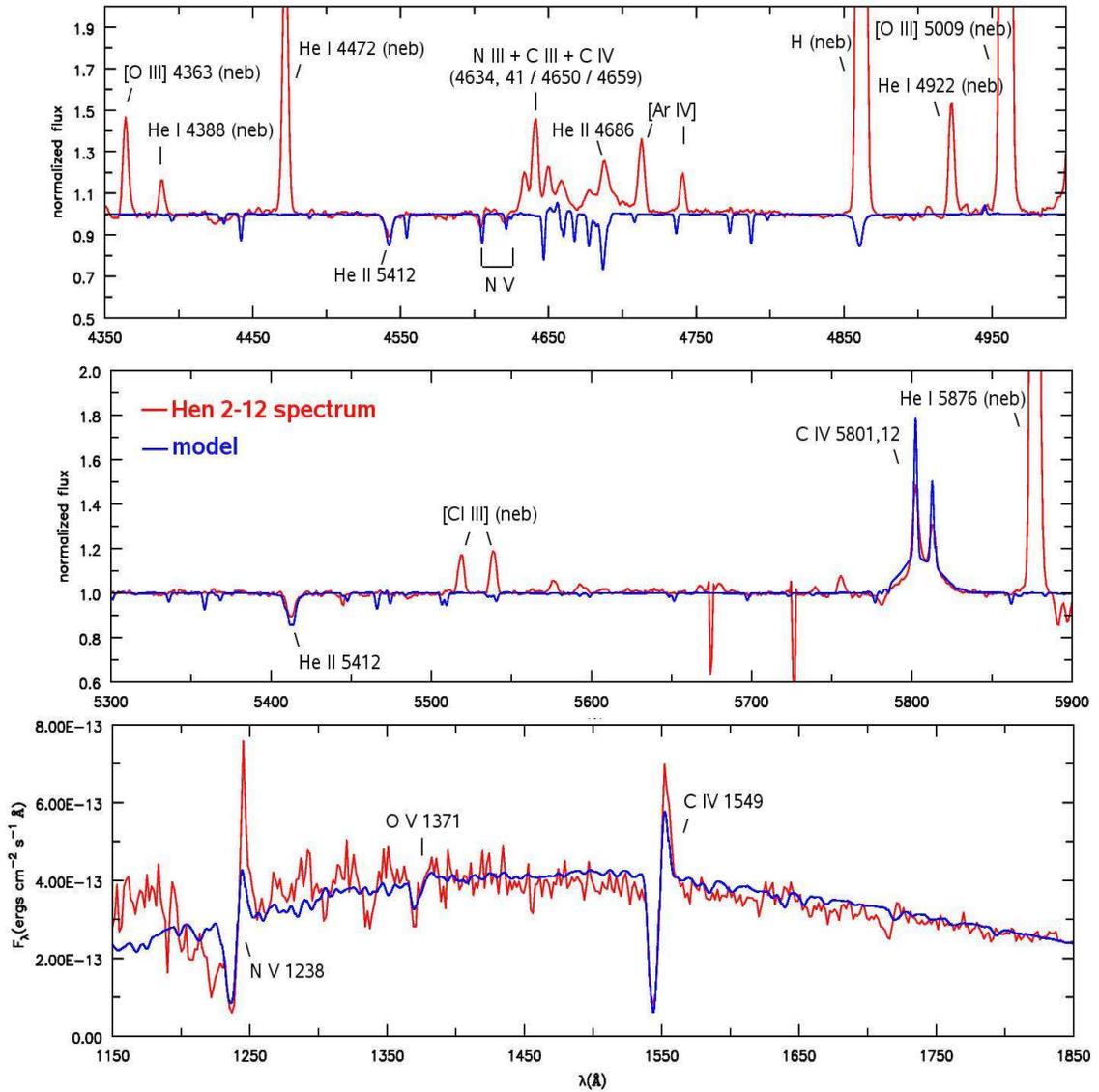} 
\caption{CMFGEN model (blue/dark grey) and observed spectra of the star Hen 2-12 (red/light grey). Optical and UV resolutions are 2\AA\, and 6\AA.} 
\label{fitcmfgen} 
\end{figure} 
 
A way to cast light in the evolutionary status of the WELS,  
is through the modeling of their spectra. In this manner,  
we can obtain their physical parameters and chemical abundances.  
Thereafter, their place in the H-R diagram could be estimated  
and a better comparison to other hydrogen deficient CSPN could   
be made. Our first efforts in this direction have been made  
by the analysis of the star Hen 2-12. We used the non LTE  
expanding atmosphere code of Hillier \& Miller (1998; CMFGEN)  
to model the spectrum of this star from the UV to the optical. 
Our fit is shown in Fig. \ref{fitcmfgen}. Several lines are  
not reproduced, since they are nebular emissions 
(they are indicated by "neb" in the figure).
 
The physical parameters derived from the fit are: 
$\dot{M} = 3.5 \times 10^{-9}M_{\sun}$ yr$^{-1}$ (clumped model, f=0.1); 
$v_\infty = 1350$ km s$^{-1}$; $R_{*} = 0.43R_{\sun}$; $T_{*} = 74$kK. 
A chemical abundance of $\beta _{C} = 70$, $\beta _{He} = 29$, 
$\beta _{O} = 0.02$ and $\beta _{N} = 0.05$, (\% by mass) 
was adopted. We can see from Fig. \ref{fitcmfgen} that the fit is not   
perfect. Some C IV absorptions are predicted in the optical, and are  
not observed. Despite these problems, the fit provides valuable   
informations. The weak absorption lines left to the blend in 4650\AA,  
are found to be due to N V. Furthermore, the lack of model emissions  
in the 4650\AA\, blend, suggest that these emissions could have a  
nebular origin. More test models are currently being computed.
 
\section{Conclusions:} 

We have investigated the UV spectra of all WELS with available IUE 
data. We have found that they can be divided in three different 
groups: (1) Strong P-Cygni profiles (mainly in C IV 1549); (2) 
Weak P-Cygni features and (3) Absence of P-Cygni profiles. 
We have derived terminal velocities for all objects presenting 
P-Cygni profiles in N V 1238 and/or C IV 1549 in a homogeneous way. 
Both the $v_\infty$ measurements and the UV characteristics observed indicate that the WELS 
are not [WC]-PG 1159 stars.

We have presented first non LTE expanding atmosphere models for the 
star Hen 2-12. Optical line fits are presented for the first time.
Although interesting results could be obtained, in order to make an 
efficient comparison to the [WR] and PG 1159 classes, and to gain insight 
to the evolutionary status of the WELS, we clearly need an analysis of a 
large sample. 
 
\acknowledgements 
Many thanks to the LOC and the hospitality of Klaus Werner and Thomas Rauch 
at the T\"ubingen meeting.

\end{document}